\documentstyle[tighten,preprint,prb,aps]{revtex}

\begin{document}

\draft
\preprint{}
\title{ Quantum chaos in a
deformable billiard:\\
Applications to quantum dots}
\author{Henrik Bruus and A.~Douglas Stone}
\address{Applied Physics, Yale University, \\P.O.Box 208284,
   New Haven, Connecticut 06520-8284}
\date{Submitted to Physical Review B15, cond-mat/9406010}
\maketitle
\begin{abstract}
We perform a detailed numerical study of energy-level
and wavefunction statistics of a deformable quantum billiard focusing
on properties relevant to semiconductor quantum dots.
We consider the family of Robnik billiards generated by simple
conformal maps of the unit disk; the shape of this family of billiards
may be varied continuously at fixed area by tuning the parameters of
the
map.  The classical dynamics of these billiards is well-understood and
this allows us to study the quantum properties of subfamilies which
span the transition from integrability to chaos
as well as families at approximately constant degree of chaoticity
(Kolmogorov entropy).  In the regime of hard chaos we find that the
statistical properties of interest are well-described by random-matrix
theory and completely insensitive to the particular shape of the dot.
However in the nearly-integrable regime non-universal behavior is
found.
Specifically, the level-width distribution is well-described by
the predicted $\chi^2$ distribution both in the presence and absence
of
magnetic flux when the system is fully chaotic; however it departs
substantially from this behavior in the mixed regime.  The chaotic
behavior
corroborates the previously predicted behavior of the peak-height
distribution for deformed quantum dots.  We also investigate the
energy-level correlation functions which are found to agree well with
the behavior calculated for quasi-zero-dimensional disordered systems.
\end{abstract}
\pacs{PACS numbers: 73.40.Gk, 05.45.+b, 72.20.My, 73.20.Dx}

\narrowtext

\section{Introduction}
The relevance of concepts from the theory of ``quantum chaos''
to mesoscopic physics has become increasingly clear
as nanostructure technology has achieved controlled
fabrication of systems smaller than both the elastic and
inelastic scattering length\cite{MesoRev}.  Quantum chaos is
the generally accepted term for properties of a quantum system
associated
with classical chaos (or the classical transition to chaos).
In the most recent high-mobility nanostructures transport is ballistic
and
the dominant scattering mechanism is the
reflection of the electrons at the boundaries of the structure
which (depending on the nature of the confining potential) may
generate
classically chaotic, mixed or integrable dynamics.
Thus they present experimental possibilities for the application and
testing of concepts from the theory of quantum chaos in condensed
matter
physics.  Disordered mesoscopic
systems also are undoubtedly chaotic classically, and recent work has
emphasized the similarity between disordered quantum systems and the
ballistic systems which have chaotic boundary scattering.  In
particular
at low temperature both type of systems exhibit sample-specific
mesoscopic fluctuations
in various physical properties as a function of external
parameters such as magnetic field.  The ballistic systems differ from
the disordered systems however in two ways.  First, as we will use the
term, a disordered system generates elastic scattering of electrons on
a scale $l$ which is short compared to typical sample dimensions $L$.
This means that transport is diffusive on a scale smaller than the
system size
and because the diffusion process is dependent on dimensionality the
statistical properties in general depend on the spatial dimension.
Chaotic ballistic systems on the other hand have no relevant transport
length smaller than the system size and thus many of their properties
are insensitive to the spatial dimensionality (they are said to be
``quasi-zero-dimensional'').  It turns out that this difference leads
to differences in the statistical properties of disordered and chaotic
systems in certain regimes of energy and
temperature\cite{Altshklov,Argaman}.
Second, there is reasonable evidence that ballistic systems may be
fabricated
with geometries and potential profiles which generate
nearly-integrable classical dynamics, thus it becomes worthwhile to
consider models which describe the transition to chaos and not just
fully chaotic dynamics.

There are two main approaches to the quantum theory of chaotic
systems.
The approach through semiclassical quantum mechanics pioneered by
Gutzwiller\cite{Gutz}, and the approach based on the theory of random
matrices
first applied to quantum chaos
by Bohigas, Giannoni and Schmit\cite{Bohigas}.  The former approach
makes a more
direct connection to the classical mechanics and has had major
successes
recently in atomic physics.  However the confining potential
in the microstructures studied experimentally is rarely well enough
known
to justify
theoretical work relying on specific classical orbits.
Instead either a wholly statistical approach, or a combination of
semiclassical and statistical ideas has been applied to quantum chaos
in
mesoscopic systems.  Three measured
physical effects which have been proposed as manifestations of
quantum chaos in mesoscopic transport are: 1) The
resistance fluctuations in GaAs quantum wires coupled strongly to
an electron cavity \cite{MesoRev,Jal90,Marcus,Keller}.  2) The weak
localization effect in the same system\cite{Keller,Barwl}.
3) The fluctuations in the
Coulomb blockade conductance peaks\cite{Meirav,Kouwenhoven}
in quantum dots weakly coupled to leads
\cite{Jal92,Jal93,Bruus,Prigo}.
The electron cavity conductance fluctuations have been described by a
combination of statistical and semiclassical theory,
and most recently certain properties have been derived from
random-matrix theory\cite{Barrmt,jalrmt}.
Whereas the Coulomb blockade peak fluctuations
in quantum dots have been described completely statistically, using
only random-matrix theory.  The quantum dot conduction experiments
are analogous to strongly resonant
scattering in atoms or nuclei for which the properties of a single
quasi-bound state can be probed.  In micron-size semiconductor quantum
dots at the
typical experimental density it is estimated that the single-particle
level-spacing (or the excitation energy to the first excited state)
$\Delta \varepsilon \sim 0.05 \; {\rm meV} \sim  500 \; {\rm mK}$ and
therefore
these systems may
be studied in the regime $kT < \Delta \varepsilon$ where indeed only a
single
quasi-bound state participates in the resonance.

In this paper we will concentrate on the statistical approach to
quantum
chaos through random-matrix theory.  This theory was
originally developed by Wigner, Dyson and others
to explain statistical properties of compound nuclear
resonances\cite{Mehta,Brody} but later it was conjectured by
Bohigas-Giannoni-Schmit \cite{Bohigas} to describe any quantum system
whose classical analogue is fully chaotic.  There now exists
substantial
numerical evidence supporting this conjecture as well as an analytic
argument due to Berry\cite{Berry}
which applies to a particular statistical property
(known as the $\Delta_3$ statistic) measuring the long-range rigidity of
the spectrum.  In RMT complex systems are
represented by ensembles of Hamiltonians with statistically-independent
matrix elements. These ensembles can be shown to have maximum
statistical
entropy (subject to a small number of constraints)\cite{Balian}.
There exist three different symmetry classes of such
ensembles characterized by the number $\beta = 1,2,4$ of
independent components of the matrix elements in the Hamiltonian,
$H$: In zero magnetic field without spin-dependent scattering
$H$ is real, $\beta=1$, and the
corresponding ensemble (invariant under orthogonal transformations) is
denoted the Gaussian orthogonal ensemble (GOE).
In non-zero magnetic field $H$ is complex, $\beta=2$, and the ensemble
(invariant under unitary transformations)
is denoted the Gaussian unitary ensemble (GUE). Finally for strong
spin-orbit scattering $H$ is quarternion real, $\beta=4$, and the
ensemble
(invariant under symplectic transformations) is denoted the Gaussian
symplectic ensemble (GSE). In this paper we shall neglect spin effects
(since spin-orbit scattering is neglibile in the ballistic
microstructures)
and hence only consider applications of the GOE or GUE.

One well-known property of these statistical ensembles is the
appearance
of strong level-repulsion due to the lack of any conserved quantum
numbers (other than the energy).  This makes the probability density
of level spacings, $s$, tend to zero as $s \to 0$ unlike the Poisson
distribution of uncorrelated random variables (which has been shown to
describe typical integrable systems\cite{BerryTabor}).
Using RMT it is possible to calculate the
distribution $P_{\beta}(s)$ in the
spectrum normalized such that the local average level spacing is one.
For RMT of two-dimensional matrices the level spacing distributions
are

\begin{eqnarray}
\label{PGOE} P_1(s) = & \makebox[2.4em][c]{$\frac{\pi}{2} s$}
\exp\left(-\frac{\pi}{4}s^2\right) & \; \; {\rm (GOE)}\\
\label{PGUE} P_2(s) = & \makebox[2.4em][c]{$\frac{32}{\pi^2} s^2 $}
\exp\left(-\frac{4}{\pi}s^2\right) & \; \; {\rm (GUE)}
\end{eqnarray}
Note that this distribution tends to zero as $s^{\beta}$ as $s \to 0$.
Although not exact for $N \times N$ random matrices, these formulae
(known as the
Wigner surmise) are excellent approximations to the exact results and
are conventionally used for comparison to statistical data.  The
appearance of the RMT spacing distributions at the classical
transition
to hard chaos has been verified in many numerical studies (although an
analytic derivation has still not been found).  Hence in this work we
will
only use the spacing distribution as a diagnostic for the
applicability
of RMT.

The theory of these random-matrix ensembles has tended to focus
on spectral statistics
and not on statistical properties of the eigenstates; however in the
application to quantum dot resonances treated below we shall find that
it is the eigenstate statistics which are most easily measured.
Very recently a great deal of progress has been
made in the study of ensembles in which the random-matrix varies as a
function of an external parameter such as magnetic flux.  Various
correlation functions of the spectra have been calculated and shown to
be universal upon rescaling\cite{Szafer,Simons}.
These correlation functions are not
easily measured in current experiments on quantum dots, but may
be accessible with some effort\cite{Sivan}.  Since such correlation
functions
are only calculated analytically
for random ensembles it is again worthwhile to test
their applicability to a given chaotic dynamical system.
We perform detailed comparisons of this type below.

Although the applicability of random-matrix theory to quantum systems
which
are classically chaotic is now reasonably well established,
most dynamical systems are neither fully chaotic
nor fully integrable, but instead have a mixed classical phase space
described topologically by the Kolmogorov-Arnold-Moser (KAM) theorem
\cite{Reichl} and related results.
The expectation is that RMT will give valid and universal results
for fully chaotic hamiltonians and will break down in some manner as
regions of stability appear in the phase-space.  It is of interest
then to look at a model for an ensemble of quantum dots which can span
the range from integrability to hard chaos, and can also test
universality in the chaotic limit.  A model of this type was introduced
by Robnik\cite{RobCM,RobQM} and generalized by Robnik and
Berry\cite{BerryRob}.
In this model the quantum dot is represented by a deformed circular well
with infinite walls; the deformation is described by
a quadratic or cubic conformal transformation.  This model has a number
of attractive features for our purposes.  1) The classical mechanics is
well-understood and has been studied in detail\cite{RobCM,Hayli}.
It has been shown that as the parameters
of the conformal transformation are varied the well can undergo a
standard
KAM transition to chaos.  In addition we show
that if the parameters are varied in a different manner a sequence of
different chaotic billiards are generated with roughly equivalent
degrees of classical chaos.  Thus one can use this set for statistical
averaging.  2) Robnik showed that a very efficient numerical algorithm
exists for obtaining a large number of eigenstates and eigenvalues for
this model.
3) Berry and Robnik showed that an Aharonov-Bohm flux may
be simply introduced into the well to break time-reversal symmetry
(without
changing the classical dynamics) and that this would only introduce
minor
changes in the numerical algorithm for solution of the Schr\"odinger
equation.
Thus the model is suitable for testing the universality of statistical
properties in the chaotic regime and their possible breakdown
in the mixed regime for both the orthogonal and unitary ensembles.
Preliminary results of this investigation of
significantly less breadth have been published elsewhere\cite{Bruus}.

The paper is organized as follows. In Sec.~II the deformable billiard
is defined in terms of a conformal map. In Sec.~III
the classical dynamics of the billiard is discussed with the emphasis
on characterizing the degree of chaos as a function of deformation. In
Sec.~IV the quantum mechanics of the
billiard is discussed, and the single-electron Schr\"{o}dinger equation
is solved with three different boundary conditions (Dirichlet,
Neumann, and general),
and the respective level statistics are calculated. In
Sec.~V using the shape as the external perturbation
the energy level correlation functions are calculated for the
normalized spectra and comparisons are made with conjectures for their
form stated in the literature. In Sec.~VI we turn to the problem of
fluctuating Coulomb blockade conductance peak heights in quantum dots.
A model of
a Coulomb blockade device is made based on the deformable billiard,
and the peak height distributions are calculated and compared to those
obtained from random-matrix theory. In Sec.~VII some experimental
implications are discussed and a new experiment
is suggested aimed at testing our
results. Finally in Sec.~VIII we present a concluding discussion.

\section{Definition of the deformable billiard}
A mathematically simple way of defining a continuous deformable family
of billiards was introduced by Robnik \cite{RobCM} and Berry and
Robnik \cite{BerryRob}. It is based on a conformal mapping of the unit
disk.  To make this paper self-contained and to facilitate
the presentation of some new developments of the method we shall
briefly describe the technique below.

As shown in Fig.~\ref{bildef} we study the simply connected domain
$\cal D$ with
a possibly irregular shaped boundary $\partial {\cal D}$ in the $uv$
plane. The open interior of the domain is denoted ${\cal D} \backslash
\partial {\cal D}$. The deformed billiard is defined by the infinitely
hard wall potential
$V(u,v)$ satisfying

\begin{equation}
\label{Vdef} V(u,v) = \left\{
\begin{array}{cll}
0 &, & \; \; (u,v) \in {\cal D} \backslash \partial {\cal D} \\
\infty &, & \; \; (u,v) \not\in {\cal D} .
\end{array} \right.
\end{equation}
Following Robnik \cite{RobQM} we define the shape of $\cal D$ by a
conformal
mapping $w$ of the unit disk $\cal C$
in the $xy$ plane to $\cal D$ in the $uv$ plane. Parts of
the treatment are more conveniently carried out by introducing the
complex coordinates $z$ and $w$:

\begin{equation}
z \equiv x +  i y \hspace{10mm} w = w(z) \equiv u(z) + i v(z).
\end{equation}
The boundary $\partial {\cal D}$ in the $w$ plane is thus given as the
image $w({\cal C})$  of the boundary of the unit disk
$\partial {\cal C}$ in the $z$ plane.
Throughout this work we study the cubic mapping introduced by
Berry and Robnik \cite{BerryRob},

\begin{equation} \label{w(z)}
w(z) = \frac{z + bz^2 + c e^{i \delta} z^3}{\sqrt{1+2b^2+3c^2}}, \; z
\in
{\cal C}, \end{equation}
where $b$, $c$, and $\delta$ are real parameters chosen such that
$|w'(z)| > 0$ for $z \in {\cal C}$.  Two sequences of
deformed billiards are show in Fig.~\ref{bilshape}. This cubic form
of $w(z)$ is the simplest conformal map resulting in a billiard
with no spatial symmetries (see Fig.~\ref{bilshape}b).  The even
simpler
quadratic map ($c=0$) generates a family of billiards with reflection
symmetry
(see Fig.~\ref{bilshape}a); Robnik and Berry pointed out that such
spatial symmetries
can prevent a magnetic flux from generating the orthogonal to unitary
transition\cite{RobBerry} hence we maintain the more general form
although
the simple quadratic case is sufficient for many of our calculations.
The deformed billiard is given by:

\begin{equation}
\begin{array}{rcl}
\label{interior}
\partial {\cal D} & = & \{w(z): \; |z| = 1 \} \\
{\cal D} \backslash \partial {\cal D} & = & \{w(z): \; |z| < 1 \}
\end{array}
\end{equation}
The
normalization in Eq.~(\ref{w(z)}) insures that the area of $\cal D$
remains $\pi$ for any value of the parameters.

For later reference we give the explicit form of the Jacobian $\left|
w'(z) \right|^2$ in the polar coordinates $(r,\theta)$ of the
$xy$ plane:

\begin{eqnarray}
\label{Jacobian}
|w'(z=re^{i\theta})|^2 = & [ & 1+4b^2r^2 + 9c^2r^4 +
6 c r^2\cos(2\theta+\delta) \nonumber \\[1mm]
&  & + 4 b r \cos(\theta)  +12 b c
r^3\cos(\theta+\delta)] \nonumber \\
&  / & (1+2b^2+3c^2). \end{eqnarray}

\section{The classical dynamics of the deformed billiard}
Robnik conducted a thorough study \cite{RobCM} of the classical
mechanics of a point mass moving freely in the deformed
billiard for the quadratic case ($c=0$, see
Fig.~\ref{bilshape}a).  More recently \cite{Hayli} Hayli et al. have
extended his results.  Thus in contrast to many previous works on
ballistic microstructures in which a discretized version of the
Schr\"odinger equation has been studied \cite{Jal92,Szafer,Simons}
in this case one has a detailed knowledge of the classical dynamics of
the relevant quantum system.  Robnik showed how as a function of $b$,
starting from $b=0$, the system evolves according to the KAM theorem
into a mixed phase-space exhibiting soft chaos and eventually to fully
developed chaos.  The main tool for determining the degree of chaos
has been
to construct the Poincar\'{e} surfaces of section for the bounce
map and to calculate the Kolmogorov entropy (we will define these
below).  The last large islands of stability in the Poincar\'{e}
section disappear around $b = 0.25$ (the value at which the billiard
ceases
to be convex) and Robnik originally conjectured
that the transition to hard chaos occured at this value\cite{RobCM}.
However recent work
has shown that very small islands of stability spawned by the
bifurcation
of the final stable two-cycle persist up to $b \approx 0.28$, but the
precise value of
b at which hard chaos sets in is not known.  However it has recently
been proven\cite{genius} that the quadratic billiard
at $b=0.50$ is fully chaotic making this only the third billiard
(along with the stadium and Sinai billiards) for which hard chaos has
been
demonstrated analytically.  In practice the islands of stability have
negligible weight for $b > 0.25$ and the statistical properties of
both the classical and quantum mechanics are consistent with fully
developed
chaos.

The Poincar\'{e} section is constructed by plotting the phase space
coordinates of the particle each time it reflects specularly from the
boundary
of the billiard. The Poincar\'{e} section is area-preserving if it
is constructed using conjugate canonical variables; a useful set
of coordinates for billiards\cite{Berryrev} is the arc length
$\sigma$,
measured from the origin to the point the particle hits the boundary,
and the tangential momentum  $\sin(\chi)$ at this  point ($\chi$
denotes the angle of incidence).
In Fig.~\ref{sos} we show Poincar\'{e} sections
corresponding to the deformation sequence of Fig.~\ref{bilshape}a. For
simplicity here we replace $\sigma$, the arc-length, by $\theta$ the
angular coordinate.  For small deformations these are approximately
proportional and since $\sigma$ is a monotonically
increasing function of $\theta$ the topology of the
surface of section in $\theta$ is the same as in $\sigma$.
(strictly speaking our use of $\theta$ introduces some small
violation of the area-preserving property which is unimportant since
we
are only using the SOS for illustrative purposes).
The surfaces of section in Fig.~\ref{sos} clearly show how the
deformed billiard gets more chaotic as the deformation is
increased from zero.
The many large stable islands at the beginning of the sequence
shrink and vanish sequentially until one obtains the featureless
surface of
section plot shown in Fig.~\ref{sos}d.
In contrast, if one starts from a deformed
billiard as in Fig.~\ref{bilshape}b and progresses through further
deformations, all the surface of section plots look
like the featureless plot in Fig.~\ref{sos}d, indicating that all of
these
billiards are close to hard chaos.

Quantitative information on the degree of chaos can be obtained by
calculating the Lyapunov exponent and the related Kolmogorov entropy
\cite{RobCM,Benettin}. Following Benettin \cite{Benettin} the Lyapunov
exponent can be estimated numerically by examining two trajectories
starting at the phase space points $\bbox{\xi}^a_0 =
\{\sigma^a_0,\sin(\chi^a_0)\}$
and $\bbox{\xi}^b_0 = \{\sigma^b_0,\sin(\chi^b_0)\}$ separated by a
small distance $d_0 = |\bbox{\xi}^b_0 - \bbox{\xi}^a_0| \approx
10^{-10}$. After one bounce the two points $\bbox{\xi}^a_1$ and
$\bbox{\xi}^b_1$ are reached, the distance $d_1$ between them is
calculated, and the reduction ratio $\beta_1 = d_0/d_1$ is formed. A
new rescaled point
$\bbox{\xi}^{b*}_1$ is defined by $\bbox{\xi}^{b*}_1 \equiv
\bbox{\xi}^a_1 + \beta_1 (\bbox{\xi}^b_1 - \bbox{\xi}^a_1)$. Starting
from $\bbox{\xi}^a_1$ and $\bbox{\xi}^{b*}_1$ a second bounce is
calculated, and from $\bbox{\xi}^a_2$ and $\bbox{\xi}^b_2$ a new
reduction ratio $\beta_2$ and a new rescaled point
$\bbox{\xi}^{b*}_2$ is formed and the process is repeated.
The Lyapunov exponent
$\Lambda$ which generally depends on the initial phase space point can
then be calculated as

\begin{equation}
\label{Lambda} \Lambda(\sigma^a_0,\sin(\chi^a_0)) =
- \lim_{n \rightarrow \infty} \left(
\frac{1}{t_n} \sum_{i=1}^{n} \ln \beta_i \right),
\end{equation}
where $t_n$ is the traversal time or the total length in real space of
the total $n$-bounce orbit.
 From the local measure of chaoticity constituted by $\Lambda$, which
depends on the initial points in phase space, a global measure the
Kolmogorov entropy $h$ can be formed:

\begin{equation}
\label{Kolmogorov}
h = \frac{1}{2 {\cal L}} \int_{-1}^{1} d(\sin(\chi)) \int_0^{{\cal L}}
d \sigma \; \Lambda(\sigma, \sin(\chi)),
\end{equation}
where $\cal L$ is the perimeter of the billiard and
$(\sigma,\sin(\chi))$ is the initial phase space point. The Kolmogorov
entropy $h$ provides a quantitative measure of the degree of chaos in
the billiard. In Fig.~\ref{Lyapunov} $h$ is shown for the deformation
sequences of Fig.~\ref{bilshape}. Note how $h$ grows monotonically for
the sequence (a), while it is fairly constant and significantly bigger
for the sequence (b). Thus the Kolmogorov entropy calculations
indicate that the deformation sequence Fig.~\ref{bilshape}b
is ``more chaotic'' than the quadratic billiard even at $b=0.50$ where
it is
known to be fully chaotic.  Of course the K-entropy is essentially an
average property of the phase-space and does not allow us to exclude
some very small regions of stability in the sequence
Fig.~\ref{bilshape}b.

\section{Quantum mechanics of the deformed billiard}
Following Berry and Robnik \cite{BerryRob} one can calculate
the quantum states of
a single particle moving in the interior of the deformed billiard
given by
Eq.~(\ref{interior}).
The billiard is threaded by an Aharonov-Bohm flux
tube of strength $\alpha \Phi_0$ through the origin of the $uv$ plane.
Here $\Phi_0$ is the flux quantum $h/e$ and $\alpha$ is a
dimensionless real number. Choosing the gauge

\begin{equation}
{\bf A}(u,v) = \frac{\alpha}{2\pi} \Phi_0 \left( \frac{\partial
f}{\partial v}, -\frac{\partial f}{\partial u}, 0 \right), \; \; {\rm
\; \;} f = \frac{1}{2} \ln \left( |z(w)|^2 \right),
\end{equation}
the Schr\"{o}dinger equation in $w$ coordinates,

\begin{equation}
\frac{1}{2m} (-i \hbar {\mbox{\boldmath $\nabla$}}\!_w + e {\bf A}(w))^2
\Psi(w) =
E \Psi(w),
\end{equation}
becomes  the
following in the polar coordinates $(r,\theta)$ of the $xy$ plane:

\begin{equation} \label{SE}
\begin{array}{lcc}
{\mbox{\boldmath $\nabla$}}^2_{r,\theta} \Psi(r,\theta)
- \frac{i 2 \alpha}{r^2} \partial_{\theta} \Psi(r,\theta)
& & \\
- \frac{\alpha^2}{r^2} \Psi(r,\theta)
+ \varepsilon |w'(re^{i\theta})|^2 \Psi(r,\theta) & = & 0,
\end{array}
\end{equation}
where now the energy $\varepsilon$ is given in units of
$\hbar^2/2mR^2$ and lengths
in units of $R$, $R$ being the radius of the circle $\cal C$. The
spectrum of the Hamiltonian is periodic in $\alpha$ with period 1.
Except for the two values $\alpha=0,1/2$ for which a real nondiagonal
representation of the Hamiltonian can be found the Hamiltonian is a
full
complex Hermitian operator. Hence GUE statistics is expected for all
$\alpha$ except for values close to 0 and $1/2$ at which GOE
statistics is
anticipated.  Below whenever we wish to focus on the GUE case
we choose the value $\alpha=1/4$ to stay as far away as
possible from the GOE values.

Thus using the fact that the deformed billiard is obtained by a
conformal
map of the unit disk we may replace the original problem on the
irregular
domain $\cal D$ by an equivalent problem on the unit disk moving under
a rather simple ``potential'' proportional to $|w'(re^{i\theta})|^2$.
In the simplest cases of Dirichlet or Neumann boundary conditions
discussed
in sections A and B below the boundary conditions of the equivalent
problem
are identical to those on the original domain.  Since a convenient
basis
which satisfies these boundary conditions is available
(products of Bessel functions, $J_{\nu}(\gamma r)$, and exponentials,
$\exp(il\theta)$) Eq.~(\ref{SE}) may be expressed as a linear equation
in that
(infinite) basis and after truncation may be solved numerically.
Typically in this study we truncate Eq.~(\ref{SE}) to the subspace of
the 1000 lowest energy Bessel solutions.
As noted by Robnik, the dependence of this linear equation on the
shape
parameters $b,c$, and $\delta$ can be factored out allowing a very
efficient
numerical method of solution as the shape is varied.
To discuss physical properties relating to isolated quantum dots it is
reasonable to consider this model with Dirichlet boundary conditions;
however we are also interested in fluctuation properties related to
conductance.  In this case the deformed dot must be coupled to leads
(typically through tunnel barriers) and more general boundary
conditions
must be imposed to describe its scattering resonances.  Hence in the
following subsections we discuss the solution of Eq.~(\ref{SE}) under
Dirichlet,
Neumann and finally general inhomogeneous boundary conditions.

\subsection{Dirichlet boundary conditions (DBC)}
Eq.~(\ref{SE}) with the Dirichlet boundary conditions (DBC)

\begin{equation}
\label{DBC} \Psi(w) = 0,\; \; {\rm for} \; w \in \partial {\cal D}.
\end{equation}
was considered in the original work of Berry and Robnik
\cite{BerryRob}.  As just noted,
a simplifying property of the conformal mapping is that if $\Psi(w)$
satisfy DBC on the domain $\cal D$, so will $\Psi(w(z))$ on the circle
$\cal C$. We therefore expand the solutions of the Schr\"{o}dinger
equation~(\ref{SE}) with DBC in terms of the
eigenstates $|\phi_{l,n}\rangle$ of the analogous Schr\"{o}dinger
equation
for the unit disk $\cal C$ also with DBC. These are given by

\begin{equation}
\label{phij}
\langle r,\theta|\phi_{l,n}\rangle = \frac{1}{\sqrt{\pi}
J'_{|l-\alpha|}(\gamma_{|l-\alpha|,n})}
J_{|l-\alpha|}(\gamma_{|l-\alpha|,n}r) e^{il\theta},
\end{equation}
where $r \le 1$, $J_{\nu}$ is the Bessel function of order
$\nu$,
$\gamma_{\nu,n}$ is the $n$th root of $J_{\nu}$, and $l$ is an
integer. The eigenvalues of $|\phi_{l,n}\rangle$ are given by

\begin{equation}
-{\mbox{\boldmath $\nabla$}}^2 |\phi_{l,n}\rangle = \gamma_{|l-\alpha|}^2
|\phi_{l,n}\rangle.
\end{equation}
Note that degeneracies occur in
the energy spectrum in case of zero AB-flux ($\alpha = 0$).

It is convenient to enumerate the basis states not according to the
quantum numbers $(l,n)$ but according to the position $j$ of the
eigenvalues
$\gamma^2$ arranged in ascending order. The eigenstates $\psi_p$ of
the Schr\"{o}dinger equation in the deformed billiard with DBC are now
written as the superposition

\begin{equation}
\label{psip}
|\psi_p\rangle = N_p \sum_{j=1}^{\infty} \frac{c^{(p)}_j}{\gamma_j}
|\phi_j\rangle,
\end{equation}
where $N_p$ is a normalization constant. The particular form of the
expansion coefficients is chosen so that the transformed
Schr\"{o}dinger equation (\ref{SE}) becomes a Hermitian eigenvalue
problem. The coefficients $c^{(p)}$ and the eigenvalues
$\varepsilon_p$ are
found from Eq.~(\ref{SE}) by insertion of the superposition
Eq.~(\ref{psip}) followed by multiplication with $\langle
r,\theta|\phi_i  \rangle$ and integration over $r$ and
$\theta$. After rearrangement the eigenvalue problem then takes the
form

\begin{equation}
\label{McEc}
M_{ij} c^{(p)}_j = \frac{1}{\varepsilon_p}c^{(p)}_j,
\end{equation}
The explicit form of the Hermitian matrix $M$ using
Eq.~(\ref{Jacobian}) for the Jacobian is the following:

\begin{eqnarray}
\label{Mij}
M_{ij} = [ \delta_{ij} & + & \delta_{l_i,l_{j-2}} 6 c e^{-i\delta}
I_{ij}^{(2)}  \nonumber \\
 & + & \delta_{l_i,l_{j-1}} (4bI_{ij}^{(1)} + 12bce^{-i\delta}
I_{ij}^{(3)}) \nonumber \\
 & + & \delta_{l_i,l_{j+0}} (8b^2 I_{ij}^{(2)} + 18 c^2 I_{ij}^{(4)})
\\
 & + & \delta_{l_i,l_{j+1}} (4bI_{ij}^{(1)} + 12bce^{+i\delta}
I_{ij}^{(3)} ) \nonumber \\
 & + & \delta_{l_i,l_{j+2}} 6ce^{+i\delta} I_{ij}^{(2)}
]/(1 + 2b^2 + 3c^2), \nonumber \end{eqnarray}
where the integrals $I_{ij}^{(h)}$ are given by

\begin{equation}
\label{Iijh}
I_{ij}^{(h)} = \frac{\int_{0}^{1} dr \: r^{h+1} J_{\nu_i}(\gamma_i r)
J_{\nu_j}(\gamma_j r)}{\gamma_i \gamma_j  J'_{\nu_i}(\gamma_i)
J'_{\nu_j}(\gamma_j)},
\end{equation}
with $\nu_i = |l_i-\alpha|$.
For given values of the parameters $b$, $c$, and $\delta$ we construct
the truncated $M$-matrix for the 1000 lowest eigenstates
$|\phi_j\rangle$. The 300 highest eigenvectors of $M$ corresponding to
the 300 lowest eigenstates of our original Hamiltonian are accurate
enough for our analysis. Typically the eigenvalues $\varepsilon_p$ are
more accurate than 0.001 times the average level spacing, and
among the 300 levels only a
few of the highest ones, where the presence of
the matrix truncation is felt most strongly,  have an accuracy as
low as around 0.1 times the average level spacing.

Once the energy spectrum is obtained for the lowest 300 eigenstates
various level statistics can be calculated and compared with the
predictions of RMT.  For specific shapes (parameter values)
this has been done previously by Berry and Robnik
\cite{BerryRob} and later by Goldberg et al.\ \cite{Goldberg}, finding
excellent agreement with e.g.\ the spacing distribution predicted by
RMT in the chaotic regime.
Since we will use a range of shapes not covered by these studies we
confirm the expectation that these results extend to the wider class
of chaotic billiards studied below.

The first step in the energy level analysis is to ``unfold'' the
spectrum.
This procedure corrects for the average variation in the density of
states
with energy so that the mean level-spacing of the unfolded spectrum is
always unity and data in different energy intervals may be merged.
For a smooth billiard obeying DBC
the Weyl formula states that the average number of states with energy
less than $\varepsilon$, $N(\varepsilon)$, is given (for high
energies) by \cite{Baltes}:

\begin{equation}
\label{WeylDBC}
N(\varepsilon) = \frac{1}{4}\varepsilon - \frac{{\cal L}}{4\pi}
\sqrt{\varepsilon} +
\frac{1}{6} ,
\end{equation}
where $\cal L$ is the perimeter of the billiard, which we calculate by
numerical integration. The unfolded eigenenergy corresponding
to the eigenenergy $\varepsilon_j$ is given by $N(\varepsilon_j)$, and
the $j$th
unfolded level spacing $s_j$ is thus

\begin{equation}
\label{xjsj}
s_j \equiv N(\varepsilon_{j+1}) - N(\varepsilon_j).
\end{equation}
After unfolding the calculated energy spectra a histogram for the
distribution of the level spacings can be constructed, an example of
which is shown in Fig.~\ref{DBChisto}. The hypothesis that the
histograms are following
the GOE distribution of Eq.~(\ref{PGOE}) or the GUE distribution
of Eq.~(\ref{PGUE}) can now be tested statistically. The quantitative
measure we use is the one-sided $\chi^2$-test, which estimates the
probability $P(\chi^2)$ for finding a $\chi^2$-value bigger than the
actual observed one.

In Fig.~\ref{DBCchi2} we show the results of the $\chi^2$-test for
the GOE and GUE distributions. We have calculated the energy spectra
for 78 different values of $\delta$ in the interval $[0,\pi]$ for
constant $b$ and $c$, both being
equal to 0.2 (the deformation sequence shown in Fig.~\ref{bilshape}b).
The billiard was threaded by an AB-flux tube of strength $\frac{1}{4}$
of a flux quantum. It is seen, as expected from
Ref.~\onlinecite{RobBerry}, that
the level distribution with statistical significance is given by the
GUE distribution in most of the parameter range. However, we also
note that for $\delta \approx 0$ and $\delta \approx \pi$ where the
billiard is almost symmetrical around the $u$-axis, the GOE
distribution
better describes the spectrum.
When comparing the results for the classical mechanics in Sec.~III
with those obtained here for the quantum mechanics, we obtain yet
another verification of the observation that the energy level
distribution of a quantum system whose classical counterpart exhibits
hard chaos follows Wigner's distribution.  Although this result
was expected, we note that our study provides a more complete test
of the validity of this statement than previous studies in that we
test a range of billiards with roughly constant Kolmogorov entropy.

\subsection{Neumann boundary conditions (NBC)}
A central issue
in this paper is to study not completely isolated but nearly
isolated systems where the electrons can escape. Therefore we have to
abandon the DBC of Eq.~(\ref{DBC}) which are relevant only for closed
systems.
As we will discuss later, a partially open system will require not just
boundary conditions but also matching conditions for solutions inside and
outside.  Such matching conditions can be expressed in terms of
any basis set for the region $\cal D$ which does not cause the
wavefunction
to vanish identically (as do DBC).
The mathematically simplest alternative choice is Neumann boundary
conditions
(NBC) for which the normal derivative vanishes everywhere on the boundary:

\begin{equation}
\label{NBC}
{\bf n} \!\cdot\! {\mbox{\boldmath $\nabla$}}\! \Psi(w) = 0, \; \; w
\in \partial
{\cal D},
\end{equation}
where ${\bf n}$ is the outward pointing normal of the boundary at the
point $w$.

Although not treated in previous work NBC maintain the simplicity of DBC
in that they are preserved by the conformal map, i.e.\ if
$\Psi(w)$ obeys Eq.~(\ref{NBC}) so will $\Psi(w(z))$ for $z \in
\partial {\cal C}$, and there exists a simply basis for the unit disk
satisfying NBC.
 We can therefore solve the Schr\"{o}dinger
equation Eq.~(\ref{SE}) with NBC following the steps in Sec.~IV-A. The
only modification is a change of the radial part of the basis
functions defined in $\cal C$ as they appear in Eq.~(\ref{phij}) to
the following (where a tilde is used to distinguish NBC from DBC):

\begin{equation}
\label{phitj}
\langle r,\theta|\tilde{\phi}_j\rangle =
\frac{\tilde{\gamma}_j/\sqrt{\pi}}{
\sqrt{\tilde{\gamma}^2_j-|l_j-\alpha|^2}
J_{|l_j-\alpha|}(\tilde{\gamma}_j)}
J_{|l_j-\alpha|}(\tilde{\gamma}_jr) e^{il_j\theta}.
\end{equation}
We have again ordered the basis set in ascending order according
to the eigenvalues $\tilde{\gamma}_j^2$, with $\tilde{\gamma}_j$ now
being a root of the
derivative $J'_{\nu_j}$ of the Bessel function $J_{\nu_j}$. The
solution $|\tilde{\phi}_p\rangle$ of Schr\"{o}dinger equation with NBC
in the
deformed billiard is written in analogy with Eq.~(\ref{psip}) as:

\begin{equation}
\label{psit}
|\tilde{\psi}_p\rangle = \tilde{N}_p \sum_{j=1}^{\infty}
\frac{\tilde{c}^{(p)}_j}{\tilde{\gamma}_j}
|\tilde{\phi}_j\rangle,
\end{equation}
The matrix equation (\ref{McEc}) becomes

\begin{equation}
\label{MtcEc}
\tilde{M}_{ij} \tilde{c}^{(p)}_j =
\frac{1}{\tilde{\varepsilon}_p}\tilde{c}^{(p)}_j,
\end{equation}
The explicit form of the Hermitian matrix $\tilde{M}$ is obtained from
Eq.~(\ref{Mij}) by substituting the integrals $I_{ij}^{(h)}$ with the
integrals $\tilde{I}_{ij}^{(h)}$ given by

\begin{equation}
\label{Itijh}
\tilde{I}_{ij}^{(h)} = \frac{\int_{0}^{1} dr \: r^{h+1}
J_{\nu_i}(\tilde{\gamma}_i r)
J_{\nu_j}(\tilde{\gamma}_j r)}{
\sqrt{\tilde{\gamma}_i^2-\nu_i^2} \sqrt{\tilde{\gamma}_j^2-\nu_j^2}
J'_{\nu_i}(\tilde{\gamma}_i)
J'_{\nu_j}(\tilde{\gamma}_j)},
\end{equation}
where $\nu_i = |l_i-\alpha|$. The energy level statistics under NBC
can now be studied by calculating the normalized energy spectrum using
numerical diagonalization of the truncated 1000 by 1000 sub-matrix of
$\tilde{M}$. The Weyl formula for the NBC case is \cite{Baltes}

\begin{equation}
\label{WeylNBC}
N(\tilde{\varepsilon}) = \frac{1}{4} \tilde{\varepsilon} + \frac{{\cal
L}}{4\pi} \sqrt{\tilde{\varepsilon}} +
\frac{1}{6},
\end{equation}
where in comparison with the DBC case Eq.~(\ref{WeylDBC}) the only
change is the sign of the second term on the right-hand side.

Although the change from DBC to NBC leads to relatively minor changes
in the method of solution, the
relation between the quantum mechanics and the classical mechanics
studied in Sec.~III becomes less clear since it is natural to
associate the hard walls in the classical problem with
DBC in the quantum problem.  Thus it is of some interest
to see if the relation between classical chaos and RMT is independent
of
this change in the boundary conditions of the quantum problem.
The histograms of the level spacing distributions for the NBC case are
calculated for the same  AB-flux and the same shape deformation
sequence as for the DBC
case, and $\chi^2$-tests for the GOE and GUE level statistics are
performed. The result is shown in Fig.~\ref{NBCchi2}, where it is seen
that again the level spacing distributions are well
described by GUE statistics, except near the special values of
$\alpha=0,1/2$.  We observe that the crossover region between GOE and
GUE
statistics near the $\delta = 0$ is wider in the NBC case than the DBC
case.  We have no physical explanation for this behavior and suspect
it
is a numerical artifact.  Nonetheless the results overall indicate
that
the correspondence between classical chaos and RMT behavior for the
spacing distribution is independent of the boundary conditions.

\subsection{General boundary conditions (GBC)}
As noted in our introduction, one proposed application of RMT to
quantum dots
relates to the peak amplitude fluctuations in resonant tunneling
through the dot.  In such a case electrons may enter and leave the dot
in certain directions through leads separated from the dot by tunnel
barriers.  Thus the problem of interest has the geometry illustrated
in
Fig.~\ref{GBCshape}.  The Schr\"odinger equation now has solutions at
all energies although resonances will still occur at an energy-spacing
comparable to that of the closed system.  For such a geometry the
physically relevant boundary conditions are Dirichlet on the portion
of
the billiard unconnected to the leads (we denote this part of the
boundary $\partial {\cal D}_1$) combined with matching conditions
at the leads.  In general one may expand the wavefunction inside the
dot in any basis which allows the matching procedure at the leads,
(i.e.\
such that the sum over basis functions is uniformly convergent).
Unfortunately Dirichlet boundary conditions at the leads do not
satisfy
this condition, thus one is forced to use a basis which satisfies
mixed boundary conditions on $\partial {\cal D}$; DBC on
$\partial {\cal D}_1$ and general inhomogenous boundary conditions on
the complement $\partial {\cal D}_2$.

\begin{equation}
\label{GBC}
\begin{array}{rccl}
g(w) \: {\bf n} \!\cdot\! {\mbox{\boldmath $\nabla$}}\! \Psi(w) +
\Psi(w) & = & 0,
& \; \;
w \in \partial {\cal D} \\
g(w) & = & 0, & \; \; w \in \partial {\cal D}_1 \\
g(w) & \neq & 0, & \; \; w \in \partial {\cal D}_2
\end{array}
\end{equation}
We refer to these as general boundary conditions, GBC.
Unlike DBC and NBC, GBC are changed by the conformal mapping;
in fact the function $g(w)$ will transform to $g(w(z))/|w'(z)|$.
However a more fundamental problem is that no simple basis set exists
on the unit disk which satisfies GBC, so no straightforward extension
of the DBC and NBC approach is possible.   Of course the problem we
are encountering occurs in other contexts (e.g. electromagnetic
waveguides)
and with significantly more effort
we are able to use existing techniques to get an approximate solution.
We employ a two-step procedure. First we find the
eigenstates of the deformed billiard obeying DBC, and then we use the
method developed by Feshbach \cite{Feshbach} to perturb the DBC at
$\partial {\cal D}_2$ with the GBC of Eq.~(\ref{GBC}). The strength of
the perturbation is given by the function $g(w)$. If $g(w)$ were zero
everywhere along $\cal D$  we recover DBC, and if $g(w)$ were infinite
everywhere along $\cal D$ we get NBC. The method we use below only
applies in the case of zero magnetic field.

Our point of departure is the eigenstates
$|\psi_p\rangle$ of Eq.~(\ref{psip}) obeying DBC. We have

\begin{equation}
\label{psistart}
{\mbox{\boldmath $\nabla$}}^2 \psi_p(w) + \varepsilon_p \psi_p(w)  = 0
\hspace*{10mm} \psi_p(S) = 0,
\end{equation}
where $S$ here and in the following denotes a point on the boundary
$\partial {\cal D}$. Our
aim is to find a solution $X_{\lambda}$ in $\cal D$ satisfying GBC:

\begin{eqnarray}
\label{chis}
{\mbox{\boldmath $\nabla$}}^2 X_{\lambda}(w) + E_{\lambda}
X_{\lambda}(w) & = & 0  \nonumber \\
g(S) \:  {\bf n} \!\cdot\! {\mbox{\boldmath $\nabla$}}\!
X_{\lambda}(S) +  X_{\lambda}(S) & = & 0.
\end{eqnarray}
We begin with the Green's function $G_{\lambda}(w|w_0)$ which
satisfies
DBC:

\begin{eqnarray}
{\mbox{\boldmath $\nabla$}}^2_0 G_{\lambda}(w|w_0) + E_{\lambda}
G_{\lambda}(w|w_0)
& = &
-\delta(w-w_0) \nonumber\\
\label{Green}
G_{\lambda}(w|S) & = & 0.
\end{eqnarray}
Eqs.~(\ref{chis}) and~(\ref{Green}) together with the use of Green's
theorem yields

\begin{eqnarray}
X_{\lambda}(w) & = & - \oint_{\partial {\cal D}} dS \: {\bf n}
\!\cdot\! {\mbox{\boldmath $\nabla$}}\!_0
G_{\lambda}(w|S) \: X_{\lambda}(S)  \nonumber \\
\label{chiInt} & = & +
\oint_{\partial {\cal D}} dS \: {\bf n} \!\cdot\! {\mbox{\boldmath
$\nabla$}}\!_0
G(w|S) \: g(S) \:   {\bf n} \!\cdot\! {\mbox{\boldmath $\nabla$}}\!
X_{\lambda}(S).
\end{eqnarray}
Taking the gradient ${\mbox{\boldmath $\nabla$}}\!_w$ of Eq.~(\ref{chiInt}) and
expanding ${\mbox{\boldmath $\nabla$}}\! X_{\lambda}$ in terms of
${\mbox{\boldmath $\nabla$}}\!
\psi_p$,

\begin{equation}
\label{gradchi}
{\mbox{\boldmath $\nabla$}}\! X_{\lambda} = \sum_p
\frac{d^{(\lambda)}_p}{\varepsilon_p}
{\mbox{\boldmath $\nabla$}}\! \psi_p,
\end{equation}
one obtains after some analysis \cite{Feshbach} the
following eigenvalue problem involving the expansion coefficients
$d^{(\lambda)}_p$ and the eigenvalues $E_{\lambda}$:

\widetext
\begin{equation}
\label{dEmatrix}
\sum_q \left\{ \left[ \frac{1}{\varepsilon_p} \delta_{p,q} +
\frac{1}{\varepsilon_p
\varepsilon_q} \oint_{\partial {\cal D}} dS \: {\bf n} \!\cdot\!
{\mbox{\boldmath $\nabla$}}
\! \psi^*_p(S) \: g(S) \:  {\bf n} \!\cdot\! {\mbox{\boldmath $\nabla$}} \!
\psi_q(S)
\right] -
\frac{1}{E_{\lambda}} \delta_{p,q}
\right\} d^{(\lambda)}_q = 0.
\end{equation}
\narrowtext
Since we are expanding the  wave function $X_{\lambda}$ obeying GBC in
a
basis set $\{ \psi_p \}$ obeying DBC some discussion of
convergence properties is necessary. The expansion in
Eq.~(\ref{gradchi})
converges only point-wise, and the convergence is slowest near the
non--DBC part $\partial {\cal D}_2$ of the boundary.
There one might expect the
occurence of ``overshooting'' (the Gibbs phenomena) of the summed
series. However, the convergence is good in a least-squares sense,
meaning that quantities involving integration over the whole domain
$\cal D$ converge well if the period of the smallest oscillation
in the series is much smaller than the extension of the region where
$g(S)$ is non-zero.  For semiclassical concepts to be relevant we
require that the leads be many wavelengths across, but we also wish
them to be small compared to the radius of the dot.
We chose to use leads with a width of about $\pi/6$ corresponding to
about five times the smallest azimuthal wave length when the basis is
truncated between 700 and 1000.
The energies $E_{\lambda}$, which is essentially the average of
$\nabla^2 X_{\lambda}$ over ${\cal D}$ is an example of a
quantity with a good least-square convergence. This convergence
improves
the smaller the strength for $g(S)$.
The DBC wavefunctions $\psi_p$ appearing
in Eq.~(\ref{gradchi}) are only known in terms of the Bessel solutions
of Eq.~(\ref{psip}).
It is thus convenient to map the line integral in Eq.~(\ref{dEmatrix})
back to the unit
disk where it becomes a sum of integrals along the perimeter
of the unit disk with $\psi_p$ and $\psi_q$ replaced by Bessel
functions.
This mapping introduces an additional factor in the integrand
$|w'(z)|^{-1}$.  A neat simplification can then be achieved by
choosing the
function $g(S)$ as

\begin{equation}
\label{g(S)}
g(S)  \equiv g(e^{i\theta}) \equiv \left\{
\begin{array}{cl}
0.1 \: |w'(e^{i\theta})|  &, \theta \in [0.5,1.0] \cup [4.5,5.0]\\
0  &, \theta \not\in [0.5,1.0] \cup [4.5,5.0]
\end{array}
\right.
\end{equation}

This choice cancels the factor $|w'(z)|^{-1}$ mentioned above and
allows
simple analytical evaluation of the line integrals.  This greatly
improves the numerical tractability of the calculation.
The strength of the perturbation $g(S)$
is chosen so that the perturbation is strong enough
to shift the individual levels of the order one mean level spacing and
such that it is weak enough to avoid substantial mixing of the
DBC-eigenstates invalidating the truncation we make at the second step
of the procedure.  In the following we show the
result of calculating the energy spectrum in the GBC case based on the
700 lowest eigenstates of the DBC case.

Once we have obtained the spectrum unfolding it requires a knowledge
of the
relevant Weyl formula since the subleading corrections depend on the
choice of
boundary conditions.  Unfortunately there does not exist an analytical
expression analogous
to Eqs.~(\ref{WeylDBC}) and~(\ref{WeylNBC}) for GBC.
However, the general theory
\cite{Baltes} implies that $N(\varepsilon)$ must have the form

\begin{equation}
\label{WeylGBC}
N(\varepsilon) = \frac{1}{4} \varepsilon + B \sqrt{\varepsilon} + C.
\end{equation}
The constants $B$ and $C$ relating to the boundary and the curvature
of
the billiard can then be estimated by numerical fitting for each
spectrum in question. We tested this procedure on the DBC and NBC
spectra and obtained the known coefficients with only a few percent
error.

In Fig.~\ref{GBCchi2} we show a comparison of $\chi^2$ tests
of the numerical spacing distribution calculated with GBC compared to
the Wigner surmise for the GOE and GUE for a range of spectra in the
deformation sequence shown in Fig.~\ref{bilshape}b for zero AB flux.
We see that GOE statistics is well-supported by the data.
Taken together the results of sections A,B and C strongly support
the conjecture that the correspondence between RMT spectral statistics
and classical chaos holds independent of the boundary conditions on
the
quantum problem.

\section{Parametric Energy Level Correlations}
Recently a great deal of theoretical work on disordered systems and
RMT has examined the statistical correlation of energy levels
when the hamiltonian varies as a function of some parameter such as
flux
\cite{Szafer,Simons,Goldberg}.
The analytic results obtained apply to disordered systems
or to RMT ensembles not generated from a microscopic hamiltonian.
Some nice numerical confirmation of the results has been obtained
for ordered but irregular systems by Szafer and Altshuler\cite{Szafer}
but
not for models in which the classical mechanics was known.
For example no numerical or analytic results are available for such
correlations in the mixed regime.
In this section we examine such correlations using the deformation
parameters
$b$ and $\delta$ as the control variable.  We use the closed billiard
with
the simple DBC discussed in Sec.~IV-A, and the deformation sequences
shown
in Fig.~\ref{bilshape}. In Fig.~\ref{bilshape}a only $b$ is changed
while
in Fig.~\ref{bilshape}b only $\delta$ varies, hence $b$ and $\delta$
will be
the control variables relevant to each sequence, which we will denote
generically as $X$. For a given sequence of parameter values the
normalized
energy spectrum is calculated. The resulting
dimensionless energy levels are denoted $\varepsilon_i(X)$.
The prediction is that for fully chaotic systems certain correlation
functions of the $\varepsilon_i(X)$ are universal upon rescaling of
$X$.
Following Ref.~\onlinecite{Simons} we define the generalized
conductance
$C(0)$ and the rescaled parameter $x$ as

\begin{equation}
\label{C0andx}
C(0) \equiv \left\langle \left( \frac{\partial
\varepsilon_i(X)}{\partial X}
\right)^2 \right\rangle \hspace{10mm} x \equiv \sqrt{C(0)} X,
\end{equation}
where $\langle \cdots \rangle$ is a statistical average over a
suitable range of energy and/or $X$.
The normalized energy level correlation function $c(x)$ defined as:

\begin{equation}
\label{cx}
c(x) \equiv \left\langle \frac{\partial \varepsilon_i(\bar{x} +
x)}{\partial \bar{x}} \frac{\partial \varepsilon_i(\bar{x})}{\partial
\bar{x}}
\right\rangle
\end{equation}
is predicted to be a universal function which differs for GOE and GUE.

The correlation function is calculated for the deformation
sequence shown in Fig.~\ref{bilshape}b with the parameter $\delta$
restricted to the interval $[0.18,2.98]$ to avoid the symmetry points
at $0$ and $\pi$. The billiard is threaded by one quarter of a flux
quantum and DBC is employed. The resulting normalized spectrum for the
50 levels between level number 225 and 274 is shown in
Fig.~\ref{espec}. Since the spectrum changes both as a function of
energy and as a function of $\delta$ the spectrum has been divided into
5 by 8 boxes each
containing 25 energy levels at 14 different values of $\delta$. Within
each box the spectrum is fairly homogeneous, and the averaging of
Eqs.~(\ref{C0andx}) and~(\ref{cx}) is performed within each
box. For each box 14 values of the pair $\{x,c(x)\}$ can thus be
calculated -- the
value of $\{0,c(0)\}$, however, is trivially $\{0,1\}$. The resulting
560
points $\{x,c(x)\}$ are then arranged in ascending order after their
first component, they are grouped in 56 groups of ten, and the average
value of
each group is calculated. The final points are displayed in
Fig.~\ref{cfunc}a. The same procedure is
repeated for the case with zero AB-flux and the resulting points are
shown in the same figure.

Fig.~\ref{cfunc}a shows an excellent agreement between the predicted
curves calculated in Refs.~\onlinecite{Szafer} and~\onlinecite{Simons}
for both the GOE and GUE cases.  Thus our results support the
conjecture
that these correlations are universal and occur when the quantum
system is classically chaotic.  We can go further however and test
whether these correlations actually coincide with classical chaos by
looking at $c(x)$ in the mixed regime.  Our results shown in
Fig.~\ref{cfunc}b
provide support for this stronger statement.  $c(x)$ in the mixed
regime
deviates strongly from its behavior for hard chaos.  In particular
a large dip appears in the correlation function which can be traced to
the average distance to the first level anti-crossing.  Unlike the
chaotic
regime where anti-crossings have large gaps, very small gap
anti-crossings
occur in the mixed regime leading to a large parametric derivative of
the energy levels.  In fact it is known \cite{RobQM}
that at such a anti-crossing in the mixed regime
two wavefunctions ``exchange identities'' with the higher energy
wavefunction
having a spatial density close the the lower-energy wavefunction which
has
just been repelled.  Thus we find that the spectral correlations are
non-universal and inconsistent with RMT in the mixed regime.

\section{Statistics of Coulomb blockade conduction peaks}
As noted above, fluctuations in the spectrum of quantum dots are not
yet easily accessible experimentally.  The most striking fluctuation
effect evident in the experimental data\cite{Meirav,Kouwenhoven,Bruus}
are fluctuations
in the peak height of Coulomb blockade resonances.  These fluctuations
reflect properties of the quasi-bound states(level-width fluctuations)
and not of the spectrum.   This contrasts with nuclear scattering
resonances in which both spectral and level-width fluctuations are
equally accessible.  The reason for this difference\cite{Jal92,Bruus}
is that the quantum
dot resonances correspond to the ground state energy of the system
with
$N,N+1,N+2,\ldots$ electrons and thus include the charging
energy $e^2/C$ associated with the addition of a particle.  Since this
charging energy is approximately constant and is typically an order of
magnitude larger than the single-particle excitation energy, $\Delta
\varepsilon$
(or more precisely the energy to the first excited state for fixed
$N$) the
observed resonances are equally-spaced to a good approximation.
In addition typically $kT >> \bar{\Gamma}$ (the mean level-width at
zero temperature) so the resonances are thermally-broadend to a width
$\sim kT$ and only their amplitude reflects the level-width
fluctuations.
The amplitude fluctuations become maximal when $kT < \Delta
\varepsilon$ and
only the ground-state contributes to the resonance.  In recent
experiments\cite{Meirav,Bruus} $\Delta \varepsilon \sim 0.5$~K so that
this
single-level regime is accessible.  The cross-over between
multiple-level
and single-level tunneling leads to unusual and fluctuating
temperature-dependences for the resonances until $kT \ll \Delta
\varepsilon$, as
was first understood by Meir et al.\cite{Meir}.
In earlier work\cite{Jal92,Jal93,Bruus}
we have developed a detailed theory of the amplitude fluctuations in
the
single-level regime assuming that RMT describes
the quasi-bound eigenstate fluctuations.  Numerical tests of the
theory
agreed well for the GOE case but not as well for the GUE
case\cite{Jal92},
and were
performed for a disordered model which was assumed to generate chaotic
classical dynamics.  Here we extend and improve these numerical tests
by
using the conformal billiard model treated above.

As before we model the quantum dot as a deformed billiard accessible
by
tunneling from leads as shown in Fig.~\ref{GBCshape}.
We neglect electron-electron interactions for the following reasons.
First they will add a charging energy which is irrelevant to the
level-width fluctuations.  Second, although the quasi-bound levels in
the presence of electron-electron interactions will surely differ from
those in its absence, we do not expect this difference to change their
statistical properties (at least in the chaotic case).  This concept
underlies the universality of RMT and is supported by experimental and
theoretical work in nuclear scattering.  For example complicated
shell-model
calculations including the residual nuclear interaction lead to
spectra
which exhibit RMT statistics \cite{Brody}.  The fact that RMT
statistics
arise in disordered or chaotic non-interacting quantum hamiltonians by
no
means imply that they are occur only when interactions are negligible.

As in nuclear physics\cite{Lane} one may relate the scattering
resonances to
the eigenstates $X_{\lambda} (u,v)$ of the dot in isolation
using R-matrix theory\cite{Jal92,Jal93,Bruus}.
In the standard approach to elastic scattering from nuclei
(for which spherical symmetry may be assumed) a linear relationship is
derived in each angular momentum channel between the scattering
wave function and its derivative at the interface between the nucleus
and free space. The coefficient of proportionality is denoted by
$R(E)$. If $M$ different decay channels exist this linear relationship
defines a matrix of coefficients known as the R-matrix for the
nuclear reaction. In our case we do not have spherical symmetry so an
angular momentum expansion is inappropriate, but we have a
simplification due to the fact that only $M$ propagating modes exist
at the Fermi energy in each of the two leads. Moreover, in general the
tunneling rate (barrier penetration factor)
will be largest for tunneling into the
lowest propagating mode, so we can in first approximation neglect all
but this mode in each lead. With this approximation
and taking the simplest case of Neumann boundary conditions for the
$X_{\lambda}$ a derivation very
similar to that used in the nuclear case yields~\cite{Jal93} a $2
\times 2$ $R$-matrix of the form

\begin{equation} \label{Rmn}
R_{mn}(E) = - \sum_{\lambda=1}^{\infty}
\frac{\gamma_{\lambda}^m\gamma_{\lambda}^n}{E-E_{\lambda}},
\end{equation}
where now the index $m\; (= r,l)$ simply denotes the
left or right leads, and the quantities $\gamma_{\lambda}^m$ and
$E_{\lambda}$
are determined by the solutions of the Schr\"{o}dinger equation
Eq.~(\ref{chis}).

\begin{equation} \label{gl}
\gamma_{\lambda}^{m} = \sqrt{P_{\lambda}} \int_{-W/2}^{+W/2} d\eta_m\:
Y(\eta_m)
X_{\lambda}(\xi_m^0,\eta_m),
\end{equation}
where $P_{\lambda}$ is the barrier penetration factor, $\xi_m$
denotes the direction parallel to lead $m$ and $\eta$ that
perpendicular,
$Y(\eta_m)$ is the transverse wave function in the leads (which have
width $W$), and $\xi_m^0$ is the position of the inner
edge of tunnel barrier $m$. In deriving this expression we assumed
that the tunnel barrier is approximately uniform in the transverse
direction.
If we were trying to derive information about specific resonances of
this system from the R-matrix then we would need to use a more general
R-matrix
than Eq.~(\ref{Rmn}) derived for the GBC relevant to our problem.
However we
are only interested in statistical properties of the R-matrix and our
results above indicate strongly that the
statistical properties are independent of the boundary conditions
imposed
to a good approximation (as long as they allow non-vanishing
wavefunctions
at the tunnel barriers).  Therefore we use the more convenient NBC for
the calculations below.

An exact non-linear relationship exists between the S-matrix and the
R-matrix\cite{Jal93};
however a particularly useful feature of this formulation is that
as $E$ approaches a particular $E_{\lambda}$, the term in
Eq.~(\ref{Rmn}) containing the
corresponding $X_{\lambda}$ will dominate (as long as coupling to the
leads
is weak) and all other terms may be neglected.
In this approximation the relationship between the $S$ and $R$ matrices
simplifies to yield the
Breit-Wigner formula for the resonance line-shape under very general
conditions. Moreover the level-width which appears in this expression
is simply

\begin{equation} \label{Gl}
\Gamma_{\lambda}^{l,r} = \frac{\hbar^2k}{m} |\gamma_{\lambda}^{l,r}|^2
\equiv
\frac{\hbar^2k}{m} P_{\lambda}^{l,r} |\tilde{\gamma}_{\lambda}^{l,r}|^2 ,
\end{equation} where we have defined the reduced width
$\tilde{\gamma}_{\lambda}$.  This reduced width
will fluctuate from level to level (and for different lead positions
for a given level) due to the complicated spatial structure of the
chaotic
eigenfunctions $X_{\lambda}$ (see Fig.~8).  For example,
if there happens to be a nodal
line near the position of a given lead then the width associated with
that
level and that lead will fluctuate down. (Note that in the absence of
spatial
symmetry another lead attached a few wavelengths away
will give a completely uncorrelated width for the same level).

In order to relate the level-widths to the experimentally-observed peak
amplitudes we use the expression for the peak height
$g_{max}$ derived by Beenakker\cite{Been91}
for the single-level regime (assuming $kT \gg \bar{\Gamma}$),

\begin{equation} \label{gmax}
g_{max} = \frac{e^2/h}{4\pi kT }\frac{\Gamma^l_{\lambda}
\Gamma^r_{\lambda}}{(\Gamma^l_{\lambda} + \Gamma^r_{\lambda})}
= \frac{e^2}{4\pi h}\frac{\bar{\Gamma}}{kT} \alpha_{\lambda}
\end{equation}
where $\Gamma_{\lambda} = \Gamma^l_{\lambda} + \Gamma^r_{\lambda}$ is
the total
decay width for level
$\lambda$ and $\Gamma^l_{\lambda},\Gamma^r_{\lambda}$ are the partial
decay widths into the right and left leads.
The factor $\alpha_{\lambda}$ in Eq.~(\ref{gmax}) is a dimensionless
measure of
the area under the $T =0$ $\lambda$-resonance, hence the observed
amplitude fluctuations reflect the fluctuations in these areas.
In this  discussion we will only treat the case of equal barriers on
each side
of the dot and hence $P_{\lambda}(E)$ will be the same on the left and
right. This
means that the {\em average} decay widths to the left and right are
equal, $\bar{\Gamma}^l = \bar{\Gamma}^r = \bar{\Gamma}/2$ and we can
express $\alpha_{\lambda}$ as

\begin{equation} \label{al}
\alpha_{\lambda} = \frac{\Gamma_{\lambda}^l
\Gamma_{\lambda}^r}{\bar{\Gamma}(\Gamma_{\lambda}^l+\Gamma_{\lambda}^r)}
= \frac{|\tilde{\gamma}_{\lambda}^l|^2
|\tilde{\gamma}_{\lambda}^r|^2
}{(|\tilde{\gamma}_{\lambda}^l|^2+|\tilde{\gamma}_{\lambda}^r|^2)}.
\end{equation}
The statistics of the peak amplitude fluctuations then follow from those
of the reduced partial widths $|\tilde{\gamma}_{\lambda}|^2$ using Eqs.
(\ref{gmax}) and (\ref{al}).

If we assume that the resonance wave functions $X_{\lambda}$ are
described by
the GOE when time-reversal symmetry is
present ($B=0$) and by the GUE when time-reversal symmetry is broken
(sufficiently high magnetic field), then the
distribution of partial widths $\Gamma_{\lambda}$ should be a
$\chi^2_{\nu}$ distribution with the degrees of freedom
$\nu=1$ and $\nu=2$ respectively \cite{Jal92}.
This distribution should be universal
in the chaotic regime, i.e.\ two different shapes both of which generate
chaotic classical dynamics should have the same distribution of level
widths (even though the individual levels are quite different).
However if the system approaches integrability then
non-universal distributions differing from $\chi^2$ should arise.
Precisely this behavior is confirmed by our numerical calculations of
the partial width distribution for the conformal billiard model as seen
in Fig.~\ref{Gdist}, where the fully chaotic GOE models (c) and (d)
fits a $\chi^2_{\nu=1}$ distribution, the fully
chaotic GUE models (e) and (f) fits a $\chi^2_{\nu=2}$ distribution,
while neither the regular model (a) nor the soft-chaotic model (b)
fits a $\chi^2_{\nu}$ distribution.
A more extensive demonstration of the universality in the chaotic
regime is
seen in Fig.~\ref{nusequence}, where we plot for each value of the shape
parameter $\delta$ in the deformation sequence of Fig.~\ref{bilshape}b
the value of $\nu$ yielding the best
$\chi^2$ fit of a $\chi^2_{\nu}$ distribution to the numerically
calculated histograms both without and with an AB-flux.
It is seen that in all cases the value of $\nu$ fluctuates around the
expected line $\nu = 1$ (zero AB-flux) and
$\nu = 2$ (non-zero AB-flux).

Having confirmed that random-matrix theory works well in the chaotic
regime, one can derive \cite{Jal92} from Eq.~(\ref{al}) the
probability density ${\cal P}_{\nu}(\alpha)$ where $\nu=2$
for the orthogonal case and $\nu=4$ for the unitary case.  One finds
\begin{eqnarray}
\label{calP2}
{\cal P}_{2}(\alpha) &=& \sqrt{2/ \pi \alpha}
\; e^{-2 \alpha} \\
\label{calP4}
{\cal P}_{4}(\alpha) &=& 4 \alpha [K_0(2 \alpha) + K_1(2 \alpha)]
\; e^{-2 \alpha}, \end{eqnarray}
where $K_n$ are the modified Bessel functions of the second kind.
${\cal P}_2$ and ${\cal P}_4$ are plotted in Fig.~\ref{P2P4} where
they are compared
to numerical data obtained by evaluating $\alpha$ in Eq.~(\ref{al})
for the
300 lowest wave functions of the conformal model for NBC in both
the GOE and the GUE case.
The time-reversal symmetry-breaking needed to study the GUE case
is achieved by adding an AB-flux of one quarter of a flux quantum. In
contrast to the results in Ref.~\onlinecite{Jal92} we find excellent
agreement between random-matrix theory and numerical calculation in
both the GOE and the GUE case.
Note the substantial suppression of small peak amplitudes caused by
breaking
time-reversal symmetry. This reduces substantially the variance of
$\alpha$, from
Eqs.~(\ref{calP2}) and~(\ref{calP4}) one finds
$\Delta \alpha_{4}^{2}/ \Delta \alpha_{2}^{2} = 32/45 \approx 0.71$.

The effect of a TR-symmetry breaking magnetic field on the
distribution and
its moments provides perhaps the simplest experimental test of our
theory.  However if the suppression of the amplitude fluctuations due
to
time-reversal symmetry breaking is to be cleanly observable then
the magnetic field necessary to
induce the GOE-GUE transition must be small compared to that needed
for Landau level formation.  Landau level formation strongly
suppresses the fluctuations\cite{Jalunpub};
the classical analogue of this effect is the suppression of chaos by
the
formation of stable skipping orbits\cite{edge}.
We estimate the magnetic field scale for TR-symmetry breaking by
adapting
an argument first put forward (to our knowledge) by Berry and Robnik
\cite{BerryRob}.  TR-symmetry is broken first not by the dynamic
effect
of the field but
by its effect on the phase of the wavefunctions (essentially the
Aharonov-Bohm effect).  Therefore in estimating the TR-symmetry
breaking
scale we may neglect the dynamic effect of the field entirely.
Gutzwiller's trace formula\cite{Gutz}
implies that structure in the spectrum on the
scale of the level spacing $\Delta \varepsilon$ arises from periodic
orbits
of period $T \approx \hbar/\Delta \varepsilon$.  A magnetic field will
change the
action (in units of $\hbar$) of such orbits by $BA_T/(h/e)$, where
$A_T$ is
the area enclosed by the periodic orbit of period $T$ in the chaotic
case.  The time-reversed orbit will of course enclose area $-A_T$ and
their
relative phases will be shifted by order unity when
$BA_T \approx (h/e)$, breaking TR-symmetry.  Thus the critical field
$B_c$ for
TR-symmetry breaking is given by $B_c \sim (h/e)/A_T$ and one need
only
estimate $A_T$.  Berry and Robnik\cite{BerryRob} treated the case of
an A-B flux as above and then evaluated the mean-squared winding
number for such orbits in the chaotic limit.
Their results can be extrapolated to a
uniform field simply by assuming a typical (positive or negative) area
of order $A$ (the area of the dot) is enclosed with each circuit.
With this modest assumption the TR-symmetry breaking flux
\begin{equation}
\label{Bc}
\Phi_c = B_cA
\sim [ \Delta \varepsilon \sqrt{A}/ \hbar v_f ]^{1/2}(h/e).
\end{equation}
Although in the experiments the dot is not isolated as assumed in this
argument, the condition $\bar{\Gamma} \ll \Delta \varepsilon$ insures
that
electrons remain in the dot long enough for the argument to still
apply.
The ratio of this
field to the field at which the cyclotron radius becomes of order
the radius of the dot scales like $N^{-3/4}$,
where N is the number of electrons;
so in dots containing a few hundred electrons TR-symmetry breaking
occurs
at a field one to two orders of magnitude smaller than that needed for
edge-state formation.  In the experimental systems of interest this
corresponds to a field of order a few times 10~mT.
Thus the statistical effect of time-reversal symmetry breaking
predicted by our theory should be observable experimentally

\section{Optimal experimental setup}
To obtain the most direct experimental verification of the theoretical
results
presented here and in earlier work it is desirable to fabricate a
quantum
dot with a variable shape.  In current experimental systems only a few
tens
of CB peaks are measured in a given dot and these are superimposed on
a significant background which complicates the comparison of the
height
of widely-separated peaks.  Thus obtaining reasonable statistics (e.g.
roughly one hundred peak amplitudes) is very difficult.
It is possible to use the magnetic field itself as an external control
parameter causing peak amplitude fluctuations \cite{Bruus}, but this
only
allows one to collect statistics for the GUE case.
Therefore we suggest an optimal heterostructure for tests of RMT
statistics
would consist of a dot formed by multiple teeth-like gates (see
Fig.~\ref{jaws}). By changing the voltages slightly on the inner gates
of this
structure it should be possible
to realize a whole range of shapes without affecting the region near
the quantum point contacts defining the tunnel barriers to the
surrounding two-dimensional electon gas. It would then be possible to
follow a given peak as a function of shape and collect statistics
without the complication of background variation, both in the presence
and absence of a magnetic field.  Since the magnetic field would not be
the control parameter it would also be possible to map out the GOE
to GUE transition.  This concept is illustrated by the calculations for
the deformed billiard shown in Fig.~16 where we plot the peak
amplitude
$\alpha$ for two given levels as function of the shape parameter
$\delta$.
A particularly simple quantity to extract from such experiments is
the correlation function for the peak amplitude as the shape is varied.
Although we are not able to make quantitative estimates at this point
our results suggest that for dots containing a few hundred electrons
(as in current experiments on micron-size devices) quite small
voltages would be needed to rearrange the wavefunctions of the
highest levels and hence decorrelate the peak amplitudes.
In the single-level regime ($kT < \Delta \varepsilon$) the theory
predicts that
the voltage scale should be independent of temperature; a prediction
which
could be experimentally tested and (if confirmed) would provide
support
for the basic model.

\section{Discussion and conclusion}
In this paper we have studied the statistical properties of deformable
billiards in the mixed and fully chaotic limits,
with applications suggested to quantum dot systems.
We have presented numerical
results based on the continuous family of conformal
quantum billiards introduced by Robnik for which efficient numerical
solution of the Schr\"{o}dinger equation is possible and for which
a rather complete characterization of the classical mechanics exists,
including the existence of a KAM transition.

This model system has provided us with a unifying framework within
which we discussed several aspects of quantum chaos in billiards.
Our calculations demonstrate the universal behavior of the system
once chaos is sufficiently strong (hard chaos) in which case
random-matrix theory describes the quantum statistical properties
independent of the details of the billiard geometry or the boundary
conditions on the quantum problem.  The calculations also
demonstrate that in the mixed regime (soft chaos)
the behavior is not universal and the random-matrix ensembles
do not describe the system.
We have also verified the theoretical
results obtained in the literature for the parametric energy level
correlations of isolated quantum dots and the
statistics of Coulomb blockade conductance peaks
based on the assumption that RMT or
the  supersymmetric $\sigma$ model applies to chaotic billiards
thereby adding credibility to these assumptions.
Finally, we have proposed an experimental setup which is particularly
well suited for tests of the theoretical predictions in this work.

\section*{Acknowledgements}
We would like to thank R.~Torsten Clay for taking part in the
preliminary calculations of the classical properties in the
billiards and B.~Simons for helpful discussions and for
allowing us to use the numerical correlation functions for disordered
systems
displayed in Fig.~\ref{cfunc}.   We thank Marko Robnik for helpful
discussions as well.  This work is partially supported by ARO grant
DAAH04-93-G-0009.   H.B.\ is grateful for support by grant no.~11-0271
from the Danish Natural Science Research Council.


\begin{figure}
\caption{\label{bildef}
The deformed billiard $\cal D$ is the image of the unit disk
$\cal C$ under the mapping $w(z)$, which is conformal ($w'(z) \neq 0,
\forall z \in {\cal C}$) and area preserving ($\int_{\cal C} |w'(z)|^2
dz = \pi$).}
\end{figure}

\begin{figure}
\caption{\label{bilshape}
Two sequences, (a) and (b), of $\partial {\cal D}$ as a
function of changing parameters. Sequence (a) was studied by Robnik
\protect\cite{RobCM}. Sequence (b) is studied in this paper. For the very
last shape in each sequence $w(z)$ is in fact not conformal. In both
of these extreme cases $w'(z) = 0$ just at the point $z=-1$.}
\end{figure}

\begin{figure}
\caption{\label{sos} Four Poincar\'{e} surface of sections for the
deformation sequence shown in Fig.~\protect\ref{bilshape}a.}
\end{figure}

\begin{figure}
\caption{\label{Lyapunov}
The Kolmogorov entropy plotted versus the normalized deformation
parameter $X$ for the two deformation sequences (a) and (b) in
Fig.~\protect\ref{bilshape}. In (a) $X=b$, and in (b) $X=\delta$.}
\end{figure}

\begin{figure}
\caption{\label{DBChisto}
The level spacing histogram for the Africa billiard $b=c=0.2$ and
$\delta=\pi/3$ with an AB-flux of $\protect\frac{1}{4} \Phi_0$ and
with
DBC. A $\chi^2$-test supports the hypothesis of GUE level statistics
(the full line) ($P(\chi^2)=0.68$), whereas the hypothesis of GOE
level statistics (the dotted line) is
rejected ($P(\chi^2)=0.0002$).}
\end{figure}

\begin{figure}
\caption{\label{DBCchi2}
The $\chi^2$ test for the deformation sequence shown in
Fig.~\protect\ref{bilshape}b in the case of DBC and with an AB-flux of
$\frac{1}{4} \Phi_0$. The full line $(\ast)$ is the test for GUE
statistics
while the dashed line $(\circ)$ is for GOE statistics.
The horizontal line is the (logarithm of the) mean value 1/2 around
which
a successful $\chi^2$ -test ought to fluctuate. Note the cross--over
between
the two statistics near the symmetry points $\delta =0$ and $\pi$.}
\end{figure}

\begin{figure}
\caption{\label{NBCchi2}
The $\chi^2$ test for the deformation sequence shown in
Fig.~\protect\ref{bilshape}b in the case of NBC and with an AB-flux of
$\frac{1}{4} \Phi_0$. The full line $(\ast)$ is the test for GUE
statistics
while the dashed line $(\circ)$ is for GOE statistics.
The horizontal line is the (logarithm of the) mean value 1/2 around
which
a successful $\chi^2$ -test ought to fluctuate.
Note the rather wide
cross over region between
the two statistics near the symmetry point $\delta =0$.}
\end{figure}

\begin{figure}
\caption{\label{GBCshape}
A deformed billiard with two leads attached. The leads can either be
completely open or they can connect to the billiard through a tunnel
barrier. The natural boundary conditions in this case are GBC. The
wavy lines inside the billiard is the nodal structure of an eigenstate
as briefly discussed in Sec.~VI.}
\end{figure}

\begin{figure}
\caption{\label{GBCchi2}
The $\chi^2$ test for the deformation sequence shown in
Fig.~\protect\ref{bilshape}b in the case of GBC and zero AB-flux.
The full line  $(\ast)$ is the test for GUE statistics
while the dashed line  $(\circ)$ is for GOE statistics.
The horizontal line is the (logarithm of the) mean value 1/2 around
which
a successful $\chi^2$ test ought to fluctuate.}
\end{figure}

\begin{figure}
\caption{\label{espec}
Part of the normalized energy spectrum (levels 225 to 274) of the
deformation sequence
Fig.~\protect\ref{bilshape}b with an AB-flux of one quarter of a flux
quantum and with DBC.}
\end{figure}

\begin{figure}
\caption{\label{cfunc}
In (a) is shown the numerical calculated energy level correlation
function $c(x)$ for the fully chaotic
deformation sequence Fig.~\protect\ref{bilshape}b
both with
($\ast$) and without ($\circ$) an AB-flux. The full (dotted)
curve is the universal correlation function calculated with (with out)
an AB-flux for disordered systems in Ref.\protect\onlinecite{Simons}.
In (b) $c(x)$ is shown with zero AB-flux in both the soft chaos regime
($\bigtriangleup$) of the first half of
the deformation sequence Fig.~\protect\ref{bilshape}a and the hard
chaos regime ($\circ$) repeated from (a). Note the strong dip for
small values of $x$ in the case of soft chaos.}
\end{figure}

\begin{figure}
\caption{\label{Gdist}
The distribution of the partial decay width,
$\Gamma_{\lambda}$, for six different deformed quantum billiards.
Shown are histograms representing our numerical results,
$\chi^2_{\nu}$ distributions (full curves)
serving as a guide for the eye ($\nu =$ 1.0 in (a)-(d) and 2.0 in
(e)-(f)),
and as inserts the particular shapes of the dot. In (a), (b),
and (c) we used the simple quadratic map with $c=0$ and with $b =$
0.00, 0.14, and 0.39
respectively. In (d), (e), and (f) we used the cubic map with
$b = c = 0.2$ and $\delta=$ $\pi/3$, $2\pi/3$, and  $\pi/3$ . In (a)
the
model is integrable, in (b)
it is nearly integrable, and in (c)-(f) it is fully chaotic. There is
no
AB-flux in (a)-(d) and one quarter of a flux quantum in (e) and (f).}
\end{figure}

\begin{figure}
\caption{\label{nusequence}
The values of $\nu$ obtained from $\chi^2$ fits of a $\chi^2_{\nu}$
distribution to the numerically calculated histograms of the
distribution of the partial decay widths $\Gamma_{\lambda}$ for the
entire deformation sequence in Fig.~\protect\ref{bilshape}b with zero
AB-flux ($\circ$) and with an AB-flux of 1/4$\Phi_0$ ($\ast$).}
\end{figure}

\begin{figure}
\caption{\label{P2P4}
Predicted distribution of peak amplitudes $\alpha$ in the presence
(a) and in the absence (b) of time-reversal symmetry, compared to
the numerically generated amplitude distribution obtained with the
shape parameters $b=c=0.2$ and $\delta = \pi/3$.
}
\end{figure}

\begin{figure}
\caption{\label{jaws}
A top view of a gated heterostructure. The gate is splitted
into twelwe teeths. The two pairs at the ends form quantum point
contacts
leading to the surrounding two-dimensional electron gas. The remaining
eight teeths define the shape of the quantum dot. In situation (a)
all the interior gates have the same voltage. In (b) one gate has a
slightly higher voltage and another slightly lower thereby deforming
the dot while maintaining its area. The regions close to the quantum
point contacts are essentially unaffected by this action.}
\end{figure}

\begin{figure}
\caption{ \label{alphacor}
The variation in the peak amplitude $\alpha$ for two
particular levels as
a function of shape deformation $\delta$.}
\end{figure}


\begin{thebibliography}{99}

\bibitem{MesoRev} For review of the physics of semiconducting
mesoscopic
devices see C.W.J.~Beenakker and
H.~van~Houten, Solid State Phys.\ {\bf 44}, 1 (1991).  Reviews of some
aspects of quantum chaos in nanostructures can be found in Chaos
{\bf 3}, (1993), articles by Marcus et al, Lin et al.\ and Baranger et
al.

\bibitem{Altshklov} B.L Altshuler and B.I. Shklovskii, Zh.\ Eksp.\
Teor.\ Fiz .
{\bf 91}, 220 (1986) [Sov. Phys.-JETP {\bf 64}, 127 (1986)].

\bibitem{Argaman} N. Argaman, Y. Imry and U. Smilansky, Phys. Rev. B
{\bf 47},
4440 (1993).

\bibitem{Gutz} For a recent review see M.~C.~Gutzwiller, {\em Chaos in
Classical and Quantum Mechanics}, Springer Verlag (New York, 1990)

\bibitem{Bohigas} O.~Bohigas, M.-J.~Giannoni, and C.~Schmit, Phys.\
Rev.\ Lett.\ {\bf 52}, 1 (1984).

\bibitem{Jal90} R.~A.~Jalabert, H.~U.~Baranger, and A.~D.~Stone,
Phys.\ Rev.\ Lett.\ {\bf 65}, 2442 (1990).

\bibitem{Marcus} C.~M.~Marcus, A.~J.~Rimberg, R.~M.~Westervelt,
P.F.~Hopkins, and A.~C.~Gossard, Phys.\ Rev.\ Lett.\ {\bf 69}, 506
(1992).

\bibitem{Keller} M.~W.~Keller, O.~Millo, A.~Mittal, D.~E.~Prober, and
R.~N.~Sachs, Surf. Sci. {\bf 305}, 501 (1994).

\bibitem{Barwl} H.~U.~Baranger, R.~A.~Jalabert, and A.~D.~Stone,
Phys.\ Rev.\ Lett.\ {\bf 70}, 3876 (1993).

\bibitem{Meirav} U.\ Meirav, M.~A.\ Kastner and S.~J.\ Wind,
Phys.\ Rev.\ Lett.\ {\bf 65}, 771 (1990).

\bibitem{Kouwenhoven} L.~P.~Kouwenhoven et al., Z.\ Phys.\ B {\bf 85},
367 (1991).

\bibitem{Jal92} R.\ A.\ Jalabert, A.\ D.\ Stone, and Y.\ Alhassid,
Phys.\ Rev.\ Lett.\ {\bf 68}, 3468 (1992).

\bibitem{Jal93} A.D.~Stone, R.A.~Jalabert, and Y.~Alhassid, in:
Springer Series in Solid-State Sciences, Vol.\ 109, Eds.\ H.\ Fukuyama
and T.\ Ando (Springer, Berlin, 1992) p.\ 39.

\bibitem{Bruus} A.~D.~Stone and H.~Bruus, Physica B {\bf 189} 43
(1993), and Surf.\ Sci.\ {\bf 305} 490 (1994).

\bibitem{Prigo} V.~N.~Prigodin, K.~B.~Efetov, and S.~Iida, Phys.\
Rev.\ Lett.\ {\bf 71} 1230 (1993).

\bibitem{Barrmt} H.U.~Baranger and P.A.~Mello, unpublished.

\bibitem{jalrmt} C.W.J.~Beenakker, R.A.~Jalabert and J.-L.~Pichard,
unpublished.

\bibitem{Mehta} M.~L.~Mehta, {\em Random Matrices and the Statistical
Theory of Energy Levels} (Academic, New York, 1991) 2nd ed.

\bibitem{Brody} T.~A.~Brody, J.~Flores, J.~B.~French, P.~A.~Mello,
A.~Pandey, and S.~S.~M.~Wong, Rev.\ Mod.\ Phys.\ {\bf 53}, 385 (1981).

\bibitem{Berry} M.~V. Berry, Proc. Roy. Soc. Lond. A {\bf 400}, 229,
(1985).

\bibitem{Balian} R.~Balian, Nuovo Cimento {\bf 57}, 183 (1968).

\bibitem{BerryTabor} M.~V.~Berry and M.~Tabor, Proc.\ R.\ Soc.\ Lond.\
A {\bf 356}, 375 (1977).

\bibitem{Szafer} A.~Szafer and B.~L.~Altshuler, Phys.\ Rev.\ Lett.\
{\bf 70}, 587 (1993)

\bibitem{Simons} B.~D.~Simons and B.~L.~Altshuler, Phys.\ Rev.\ Lett.\
{\bf 70}, 4063 (1993) and Phys.\ Rev.\ B {\bf 48}, 5422 (1993).

\bibitem{Sivan} U.~Sivan et al. Europhys.\ Lett.\ {\bf 25}, 605 (1994).

\bibitem{Reichl} L.E.~Reichl, {\em The Transition to Chaos} (Springer
Verlag, New York, 1992).

\bibitem{RobCM} M.~Robnik, J.~Phys.~A {\bf 16}, 3971 (1983).

\bibitem{RobQM} M.~Robnik, J.~Phys.~A {\bf 17}, 1049 (1983).

\bibitem{BerryRob} M.~V.~Berry and M.~Robnik, J.~Phys.~A {\bf 19},
649 (1986).

\bibitem{Hayli} A.~Hayli, T.~Dumont, J.~Moulin-Ollagnier, and J.-M.~
Strelcyn, J.\ Phys. A {\bf 20}, 3237 (1987).

\bibitem{RobBerry} M.~Robnik and M.~V.~Berry, J.~Phys.~A {\bf 19},
669 (1986).

\bibitem{genius} R.~Markarian, Nonlinearity {\bf 6}, 819 (1993)

\bibitem{Berryrev} M.~V.~Berry, Eur.\ J.\ Phys.\ {\bf 2}, 91 (1981).

\bibitem{Benettin} G.~Benettin and J.--M.~Strelcyn, Phys.\ Rev.\ A
{\bf 17} 773 (1978).

\bibitem{Goldberg} J.~Goldberg, U.~Smilansky, M.~V.~Berry,
W.~Schweizer, G.~Wunner, and G.~Zeller, Nonlinearity {\bf 4}, 1 (1991).

\bibitem{Baltes} H.~P.~Baltes and E.~R.~Hilf {\em Spectra of Finite
Systems} (Wissenschaftsverlag, Bibliographisches Institut, Z\"{u}rich,
1976).

\bibitem{Feshbach} H.~Feshbach, Phys.\ Rev.\ {\bf 65}, 307 (1944), and
Chap.\ 9 in P.~M.~Morse and H.~Feshbach, {\em Methods of theoretical
physics} (McGraw-Hill, New York, 1953).

\bibitem{Meir} Y.~Meir, N.~Wingreen and P.~A.~Lee,
Phys.\ Rev.\ Lett.\ {\bf 66}, 3048 (1991).

\bibitem{Lane} A.M. Lane and R.G. Thomas, Rev. Mod. Phys. {\bf 30},
257 (1958).

\bibitem{Been91} C.~W.~J.~Beenakker, Phys.\ Rev.\ B {\bf 44}, 1646
(1991).

\bibitem{Jalunpub} R.A. Jalabert and A.D. Stone, unpublished.

\bibitem{edge} M.~Robnik and M.~V.~Berry, J.~Phys.~A {\bf 18},
1361 (1985).


\end{thebibliography}
\end{document}